\documentclass[twocolumn]{aastex631}

\shorttitle{Evidence for accretion bridge in the DX Cha circumbinary system}
\shortauthors{Juhász et al.}

\graphicspath{{./}{figures/}}

\usepackage{graphicx}
\usepackage{txfonts}
\usepackage{placeins}
\usepackage{float}
\usepackage{xcolor}
\usepackage{comment}
\usepackage{upgreek}

\begin{document}

    \title{Evidence for an accretion bridge in the DX Cha circumbinary system from VLTI/MATISSE observations}
    \thanks{Based on observations collected at the European Southern Observatory \\ under ESO programmes 0104.C-0782(D), 106.21Q8.003,\\ 108.22HB.001 and 110.23X2.002.}

\newcommand{\Konkoly}{Konkoly Observatory, HUN-REN Research Centre for Astronomy and Earth Sciences, Konkoly-Thege Mikl\'os \'ut 15-17, 1121 Budapest, Hungary}
\newcommand{\MPIA}{Max Planck Institute for Astronomy, K\"onigstuhl 17, D-69117 Heidelberg, Germany}
\newcommand{\ELTE}{Institute of Physics and Astronomy, ELTE E\"otv\"os Lor\'and University,  P\'azm\'any P\'eter s\'et\'any 1/A, 1117 Budapest, Hungary}
\newcommand{\Leiden}{Leiden Observatory, Leiden University, P.O. Box 9513, 2300 RA Leiden, The Netherlands}
\newcommand{\MTA}{CSFK, MTA Centre of Excellence, Konkoly-Thege Miklós út 15-17, H-1121 Budapest, Hungary}
\newcommand{\Toulouse}{Institut de Recherche en Astrophysique et Planétologie, Universit\'e de Toulouse, UT3-PS, CNRS, CNES, 9 av. du Colonel Roche, 31028 Toulouse Cedex 4, France}
\newcommand{\CdA}{Universit\'e C\^{o}te d’Azur, Observatoire de la C\^{o}te d’Azur, CNRS, Laboratoire Lagrange, France}
\newcommand{\Amsterdam}{Anton Pannekoek Institute for Astronomy, University of Amsterdam, the Netherlands}
\newcommand{\Grenoble}{Univ. Grenoble Alpes, CNRS, IPAG, 38000 Grenoble, France}
\newcommand{\NASA}{NASA Goddard Space Flight Center, Astrophysics Division, Greenbelt, MD 20771, USA}
\newcommand{\Kiel}{Institute of Theoretical Physics and Astrophysics, University of Kiel, Leibnizstr. 15, 24118 Kiel, Germany}
\newcommand{\Liege}{STAR Institute, University of Li\`ege, Li\`ege, Belgium}
\newcommand{\Paris}{AIM, CEA, CNRS, Universit\'e Paris-Saclay, Universit\'e Paris Diderot, Sorbonne Paris Cit\'e, F-91191 Gif-sur-Yvette, France}
\newcommand{\MPIR}{Max-Planck-Institut für Radioastronomie, Auf dem Hügel 69, 53121, Bonn, Germany}
\newcommand{\Radboud}{Institute for Mathematics, Astrophysics and Particle Physics, Radboud University, P.O. Box 9010, MC 62 NL-6500 GL Nijmegen, the Netherlands}

\author[0000-0003-1719-8503]{T\'\i{}mea Juh\'asz}
\affiliation{\Konkoly}
\affiliation{\ELTE}
\affiliation{\MTA}
\email{juhasz.timea@csfk.org}

\author[0000-0003-4989-575X]{J\'ozsef Varga}
\affiliation{\Konkoly}
\affiliation{\MTA}

\author[0000-0001-6015-646X]{P\'eter \'Abrah\'am}
\affiliation{\Konkoly}
\affiliation{\MTA}
\affiliation{\ELTE}

\author[0000-0001-7157-6275]{\'Agnes K\'osp\'al}
\affiliation{\Konkoly}
\affiliation{\MTA}
\affiliation{\ELTE}

\author[0000-0002-6394-8013]{Foteini Lykou}
\affiliation{\Konkoly}
\affiliation{\MTA}

\author[0000-0003-2835-1729]{Lei Chen}
\affiliation{\Konkoly}
\affiliation{\MTA}

\author[0009-0001-9360-2670]{Attila Mo\'or}
\affiliation{\Konkoly}
\affiliation{\MTA}

\author[0000-0002-4283-2185]{Fernando Cruz-S\'aenz de Miera}
\affiliation{\Toulouse}
\affiliation{\Konkoly}
\affiliation{\MTA}

\author[0000-0000-0000-0000]{Bruno Lopez}
\affiliation{\CdA}

\author[0000-0000-0000-0000]{Alexis Matter}
\affiliation{\CdA}

\author[0000-0000-0000-0000]{Roy van~Boekel}
\affiliation{\MPIA}

\author[0000-0000-0000-0000]{Michiel Hogerheijde}
\affiliation{\Leiden}
\affiliation{\Amsterdam}

\author[0000-0000-0000-0000]{Margaux Abello}
\affiliation{\CdA}

\author[0000-0000-0000-0000]{Jean-Charles Augereau}
\affiliation{\Grenoble}

\author[0000-0000-0000-0000]{Paul Boley}
\affiliation{\MPIA}

\author[0000-0000-0000-0000]{William C. Danchi}
\affiliation{\NASA}


\author[0000-0000-0000-0000]{Thomas Henning}
\affiliation{\MPIA}






\author[0000-0000-0000-0000]{Mathis Letessier}
\affiliation{\Grenoble}

\author[0000-0000-0000-0000]{Jie Ma}
\affiliation{\Grenoble}





\author[0000-0000-0000-0000]{Philippe Priolet}
\affiliation{\Grenoble}


\author[0000-0000-0000-0000]{Marten Scheuck}
\affiliation{\MPIA}




\author[0000-0000-0000-0000]{Gerd Weigelt}
\affiliation{\MPIR}

\author[0000-0000-0000-0000]{Sebastian Wolf}
\affiliation{\Kiel}



\begin{abstract}

DX Cha (HD 104237) is a spectroscopic binary consisting of a Herbig A7.5Ve-A8Ve primary star and a K3-type companion. Here we report on new $3.55$ $\upmu$m interferometric observations of this source with the Multi Aperture Mid-Infrared Spectroscopic Experiment (MATISSE) at the Very Large Telescope Interferometer (VLTI). To model the four MATISSE observations obtained between 2020 and 2023, we constructed a time-dependent interferometric model of the system, using the \texttt{oimodeler} software. The model consists of an asymmetric ring and two point sources on a Keplerian orbit.
Our best-fit model consists of a circumbinary ring with a diameter of $0.86$ au ($8.1$ mas), featuring a strong azimuthal asymmetry. We found that the position angle of the asymmetry changes tens of degrees between the MATISSE epochs. The ring is relatively narrow, with a full width at half maximum (FWHM) of $\sim$$0.13$ au ($1.23$ mas). 
The presence of circumstellar dust emission so close to the binary is unexpected, as previous hydrodynamic simulations predicted an inner disk cavity with a diameter of $\sim$$4$ au ($\sim$$37.5$ mas). Thus, we argue that the narrow envelope of material we detected is probably not a gravitationally stable circumbinary ring, but may be part of tidal accretion streamers channeling material from the inner edge of the disk toward the stars.


\end{abstract}


\section{Introduction}
\label{sec:intro}

Herbig Ae stars are young ($<10$\,Myr) A-type pre-main sequence stars with masses of ~1.5–-3 $M_\odot$. A circumstellar disk is typically present around Herbig stars, the state and evolution of which are closely related to the central star \citep{alonso2009}. The evolution of these young systems is particularly complex in the case of circumbinary disks, where the orbital motion and accretion of two central stars have profound effects on the disk \citep{2002A&A...387..550G, 2024MNRAS.528.7256T}.
The distribution of the circumstellar material can be very varied,
depending on the mass ratio and the separation of the two stars, as well as on the eccentricity of their orbits.

DX\,Chamaeleontis (DX\,Cha, HD 104237) is a spectroscopic binary which belongs to the $\sim$5\,Myr old $\epsilon$ Cha association \citep{murphy2013,dickson-vandervelde2021}. The main stellar parameters are shown in the Table \ref{tab:fix}.

\begin{table}
\caption{Stellar parameters of the DX Cha system. \textbf{References.} (1) \cite{2004ApJ...608..809G};(2) \cite{2004A&A...427..907B}; (3) \cite{2013MNRAS.430.1839G}; (4) \cite{2013MNRAS.431.3485C}; (5) \cite{2021yCat.1352....0B}. We fit the flux ratios, the other parameters are fixed in the modeling. $\star$We use $0^{\circ}$ inclination for the circumbinary ring in our modeling.}
\begin{center}
    \label{tab:fix}
    \small
    \begin{tabular}{c c c}
        \hline
        \hline
         Parameter & Value & Reference\\
        \hline
        \textbf{Primary star} & &\\
        Spectral type & A7.5Ve-A8Ve & (1)\\
        Mass & $2.2\pm0.2$~M$_{\odot}$ & (3)\\
        Temperature & $8250\pm200$ K & (4)\\
        \hline
        \textbf{Secondary star} & &\\
        Spectral type & K3 & (1)\\
        Mass & $1.4\pm0.3$~M$_{\odot}$ & (3)\\
        Temperature & $4800\pm200$ K & (4)\\
        \hline
        \textbf{Binary orbit} & &\\
        Orbital period & $19.859$ d & (2)\\
        Separation & $0.22$ au & (3)\\
        Eccentricity & $0.665$ & (2)\\
        Longitude of periastron & $216.082^{\circ}$ & (2) \\
        Longitude of the ascending node & $235^{\circ}\pm3$ & (2) \\
        Inclination$\star$ & $17^{\circ}{}^{+12}_{-9}$ & (3) \\
        \hline
        Distance & $106.5\pm0.5$~pc & (5) \\
        \hline
         
    \end{tabular}
\end{center}
\end{table}

\cite{2001A&A...365..476M} studied the circumstellar material of DX Cha and other $13$ Herbig Ae/Be stars using the Short Wavelength Spectrometer at ISO \citep{iso_sws_1996}. Their analysis suggested that DX Cha belongs to group II in their YSO classification.

DX Cha has been extensively observed with the infrared (IR) interferometric instruments of the Very Large Telescope Interferometer (VLTI), resolving its circumstellar environment at sub-au scales. \cite{2007A&A...464...55T} observed DX Cha with VLTI/AMBER \citep{petrov2007}, and they suggested that its $K$-band circumstellar emission originates in a “puffed-up” inner rim of a disk. They also concluded that the detected Br$\gamma$ emission line probably comes from a compact disk wind, launched from a region $0.2$--$0.5$ au from the star. \cite{Kraus2008} measured a continuum ring radius of $0.29$~au, also from AMBER data, that is close to the expected dust sublimation radius ($0.32$~au) for sub-micron sized silicate grains. \cite{Kraus2008} suggested that the $K$-band emission may trace the inner rim of a circumbinary dust disk.
\cite{2013MNRAS.430.1839G} found a slightly larger circumbinary disk with a radius of $\sim$$0.5$ au, also from AMBER data. They found that a fraction of the $K$-band flux is unresolved, and not accounted for by the stellar flux contributions, and hypothesized that this unresolved flux may come from compact structures inside a tidally disrupted circumbinary disk. The $H$-band VLTI/PIONIER \citep{pionier2011} data show a structure which becomes fully resolved at a baseline of $\sim$$30\ \mathrm{M}\lambda$ \citep{Lazareff2017}. Based on this, we estimate the binary and disk structure have a typical diameter of 8 mas ($\sim$$0.85$ au) in the $H$-band. The PIONIER closure phases indicate significant asymmetry.

DX Cha was observed in the $N$-band with VLTI/MIDI \citep{2003Ap&SS.286...73L} several times between 2005 and 2014. The spatial resolution of MIDI permits a study of disk structure on $\sim$$1$ au scales. The system was included in the sample of $82$ young stellar objects with archival MIDI observations analysed by \cite{2018A&A...617A..83V}\footnote{They used $104\pm3$ pc for distance of DX Cha.}, who estimated a stellar luminosity of $47\pm11$~L$_{\odot}$ from the SED. They found that the half-light radius of the circumstellar emission in the $N$-band is $0.64$~au.

\cite{2020A&A...642A.104K} studied the variability of the surface brightness distribution of the disk. They found variability on timescales of $51$ days in the PIONIER data series (up to 0.23 $\hat{=}$ 31$\sigma$), and on one day timescale in the AMBER observations (0.05 $\hat{=}$ 6.7$\sigma$). Additionally, \cite{2012ApJS..201...11K} found wavelength-independent flux changes of $\geq0.1$ mag based on four mid-infrared spectra observed with ISO/ISOPHOT and Spitzer/IRS between 1996 and 2005.

\cite{2015MNRAS.448.3545D}\footnote{They used $116\pm7$ pc for distance of DX Cha.} performed smoothed particle hydrodynamics (SPH) simulations of the disk around DX Cha. The simulation predicts that the binary would clear out a $\sim$$4$ au diameter cavity in the disk. The radius of the cavity, $\sim$$9 a$, where $a$ is the semi-major axis of the binary, may seem unusually large, as in other simulations it is usually between $2a$ and $3a$ \citep[e.g.,][]{2020A&A...641A..64H}. However, in the simulation of \cite{2015MNRAS.448.3545D}, the stars move on a highly eccentric orbit, hence the larger-than-usual cavity size.

The simulations of \cite{2015MNRAS.448.3545D} also showed accretion bridges within the cavity belonging to both stars, and that the disk precesses with a period of approximately $40$ years, and suggested that the disk precession effect could be explored with VLTI, as in four years the disk will have precessed by approximately $36^{\circ}$.

In this work, we study the disk structure of DX Cha in the inner few au region with new VLTI/MATISSE \citep{matisse2022} $L$-band (3.5\,{\micron}) data from the MATISSE Guaranteed Time Observations (GTO) program. One of our main goals is to check for the presence of the cavity predicted by the SPH simulations of \cite{2015MNRAS.448.3545D}. A second objective is to detect changes in the disk structure over time. We also aim to investigate the physical cause of the potential structural variability.

This work is structured as follows: in Section \ref{sec:data} we present the observations and the data reduction. Section \ref{sec:modeling} describes the modeling method, including modeling of the SED and interferometric data, and in Section \ref{sec:results} we show our results. In Section \ref{sec:disc} we give a discussion of our results. Finally, in Section \ref{sec:sum} we present the summary of our work.

\section{Measurements and data reduction}
\label{sec:data}

\begin{table*}
\caption{Overview of VLTI/MATISSE observations of DX Cha used in our work. $\tau_0 $ is the atmospheric coherence time. LDD {(limb darkened diameter)} is the estimated angular diameter of the calibrator. $\star$\cite{Bourges2017}.}
\begin{center}
    \label{tab:obs}
    \small
    \begin{tabular}{c c c c c c c c c c}
        \hline
        \hline
         \multicolumn{7}{c}{Target} & \multicolumn{3}{c}{Calibrator}\\
        \hline
         Date and time & Instrument mode &  L band & Seeing & $\tau_0$ & Stations & Configuration & Name & LDD$\star$& Time \\
         (UTC) & & $\Delta\lambda/\lambda$ & (\arcsec) & (ms) &  &  & & (mas)& (UTC) \\
        \hline
2020-03-22T05:54 & stand-alone & 34 & 0.54 & 7.25 & K0-G2-D0-J3 & Medium (AT)& HD 105340 & 2.22 & 05:29 \\
2021-03-11T04:45 & GRA4MAT & 34 & 0.68 & 6.49 & K0-G2-D0-J3 & Medium (AT) & HD 120404 & 2.94 & 04:18 \\
2022-01-23T07:37 & stand-alone & 34 & 0.49 & 10.40 & U1-U2-U3-U4 & UT & HD 92682 & 2.12 & 07:04 \\
2023-01-18T06:15 & GRA4MAT & 506 & 0.76 & 6.48 & A0-B2-D0-C1 & Small (AT) & HD 92305 & 4.61 & 05:46 \\
        \hline
    \end{tabular}
\end{center}
\end{table*}

We used interferometric data taken with the Multi AperTure mid-Infrared SpectroScopic Experiment (MATISSE) instrument of the  VLTI, by the MATISSE Guaranteed Time Observations (GTO) team. MATISSE is a four-telescope beam combiner, observing in the $L$ (3--4~$\upmu$m), $M$ (4.5--5.0~$\upmu$m), and $N$ (8--13~$\upmu$m) bands \citep{2022A&A...659A.192L}, and can provide different spectral resolutions, low, medium, high, and very high ($R \approx 30$, 500, 1000, and 3500, respectively) in the $L$/$M$-bands, and low or high ($R \approx 30$, 220, respectively) in the $N$-band.

In this study we used the $L$-band data which have been reduced with the Data Reduction Software package (DRS) \citep{2016SPIE.9907E..23M}. Our data, as a standard procedure, contain observations of calibrator stars with known spectra and stellar disk angular sizes. This allows us to perform proper calibration of the visibilities and closure phases, and also flux calibration of the total and correlated spectra. The diameters of the calibrator stars were taken from the JSDC catalog\footnote{http://cdsarc.u-strasbg.fr/cgi-bin/VizieR?-source=II/346} \citep{Bourges2017}.  For the flux calibration, we used the MATISSE python tools\footnote{https://github.com/Matisse-Consortium/tools} \citep[for more details, we refer to][]{2024A&A...681A..47V}.
Table \ref{tab:obs} shows the log of observations. The series also contain $N$-band data, but we did not use them, as our goal was to examine the innermost emitting region which is better traced by the $L$-band data. The $N$-band data will be the subject of a separate study.

 \begin{figure*}[!ht]
\centering
    \includegraphics[width=0.98\textwidth]{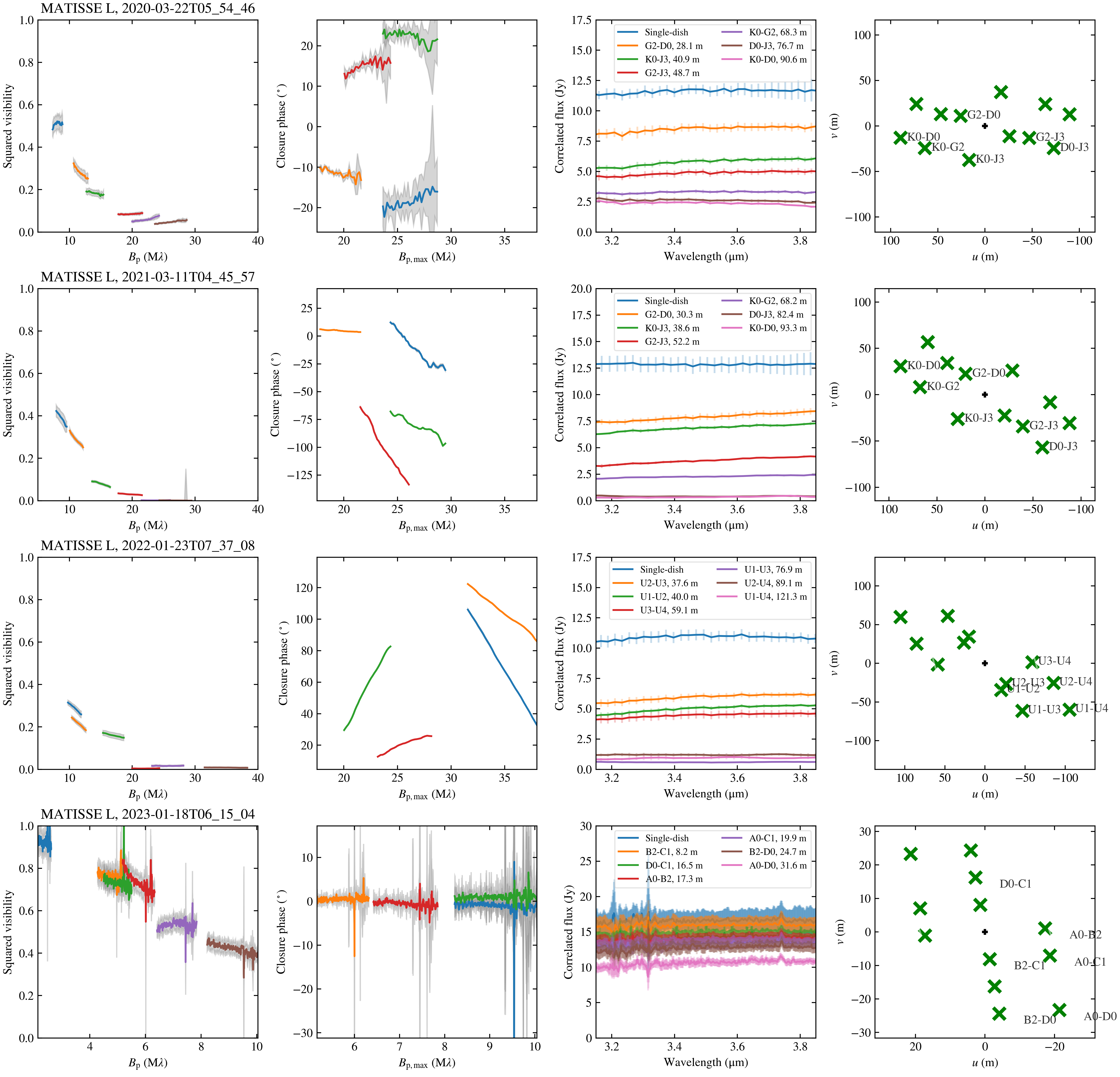}
    \caption{The calibrated MATISSE $L$-band data sets. First column: visibility as a function of the spatial frequency. Second column: closure phase as a function of the spatial frequency corresponding to the longest baselines of the triangles. Third column: correlated flux as a function of the wavelength. We also plot the single-dish spectrum, as the zero-baseline correlated spectrum. Fourth column: uv-coverage of the observations. }
    \label{fig:observations}
\end{figure*}

Plots showing the uv-coverage of the MATISSE observations along with visibilities, closure phases, and correlated fluxes can be found in the Figure \ref{fig:observations}. The measurements from 2020 and 2021 were done on the medium AT baseline configuration covering the $40$--$104$~m range in baseline length. The observation of 2022 was done on the fixed UT array with a baseline range of $46$--$130$~meters. These three data sets show closure phases significantly deviating from zero (reaching as high as $\sim$$100^\circ$), which indicates the asymmetry of the system. We note that the field of view of MATISSE on the ATs ($\mathrm{FOV}\approx600$~mas in the $L$-band) and on the UTs ($\mathrm{FOV}\approx130$~mas in the $L$-band) is significantly different. Still, that is not expected to cause any issues in the modeling because all the mid-IR emission of the source is expected to come from a region much less that $100$~mas.

The measurements from 2023 were obtained on the small AT array at $11$--$34$~m baseline lengths. At these small baselines the source is moderately resolved (visibility ($V$) drops to $\sim$$0.6$), and the closure phases barely exceed $\pm2^\circ$, therefore they do not carry much information about the asymmetry. 
For this reason, we assume that the source is not variable on the small AT baselines. In our subsequent analysis (Section \ref{sec:modeling}) we interpret the longer-baseline 2020-2022 data sets with a time-variable geometry. As the small AT data set is not expected to vary, we merge those data with each of the 2020, 2021, and 2022 data sets for the fitting, in order to have both the short and long baselines sampled.

\section{Modeling}
\label{sec:modeling}
\begin{figure}[h]
\centering
    \includegraphics[width=0.47\textwidth]{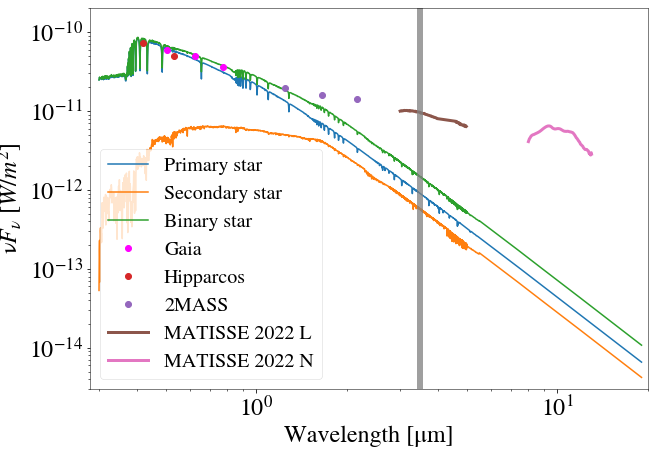}
    \caption{SED of DX Cha. The grey band shows the wavelength range we modeled.}
    \label{fig:sed}
\end{figure}

\begin{figure*}[ht]
\centering
    \includegraphics[height=0.24\textwidth]{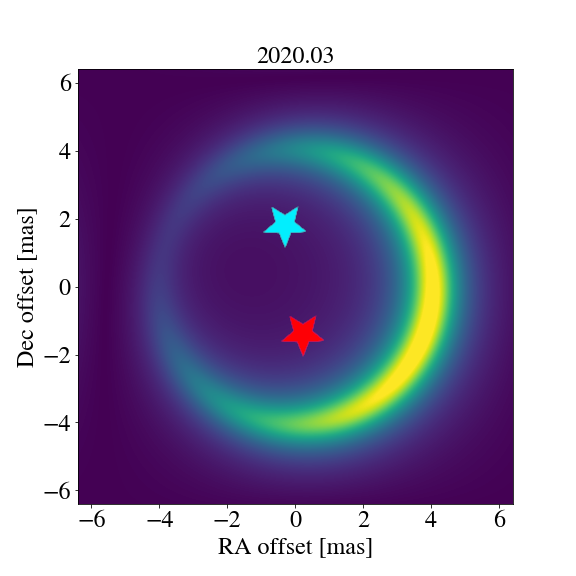}
    \includegraphics[height=0.24\textwidth]{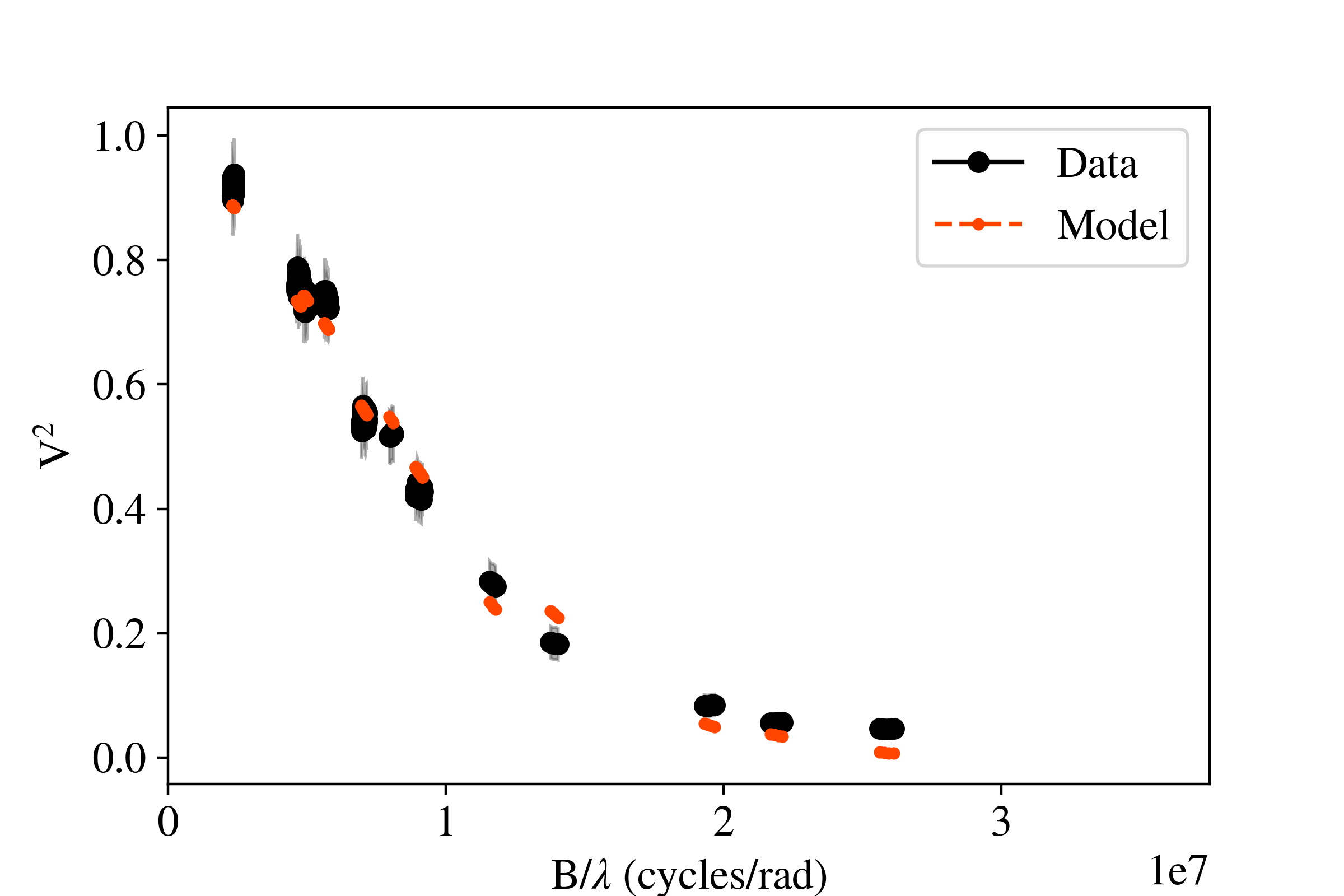}
    \includegraphics[height=0.24\textwidth]{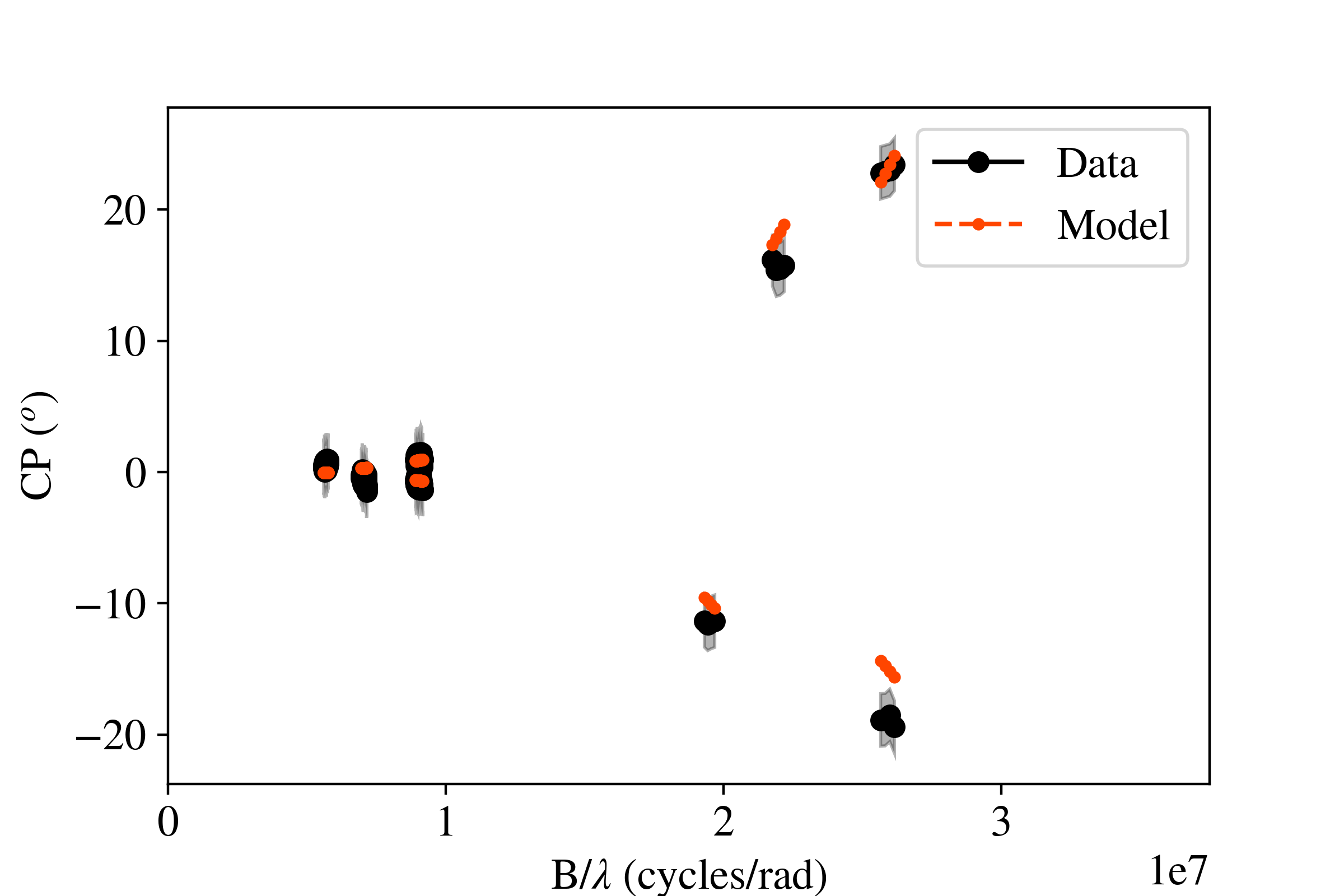}
    \includegraphics[width=0.24\textwidth]{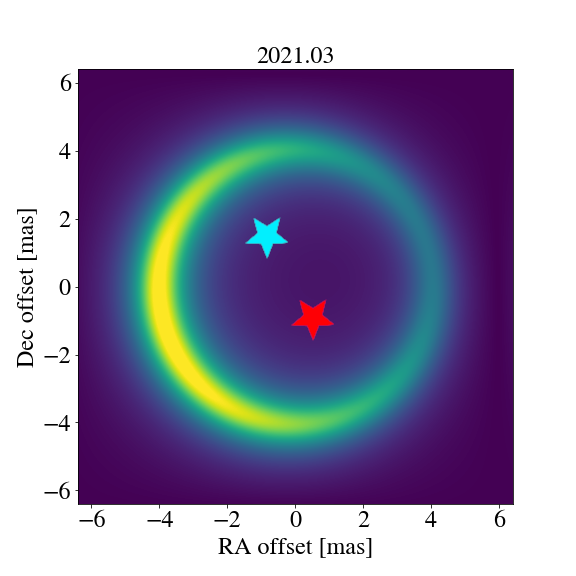}
    \includegraphics[height=0.24\textwidth]{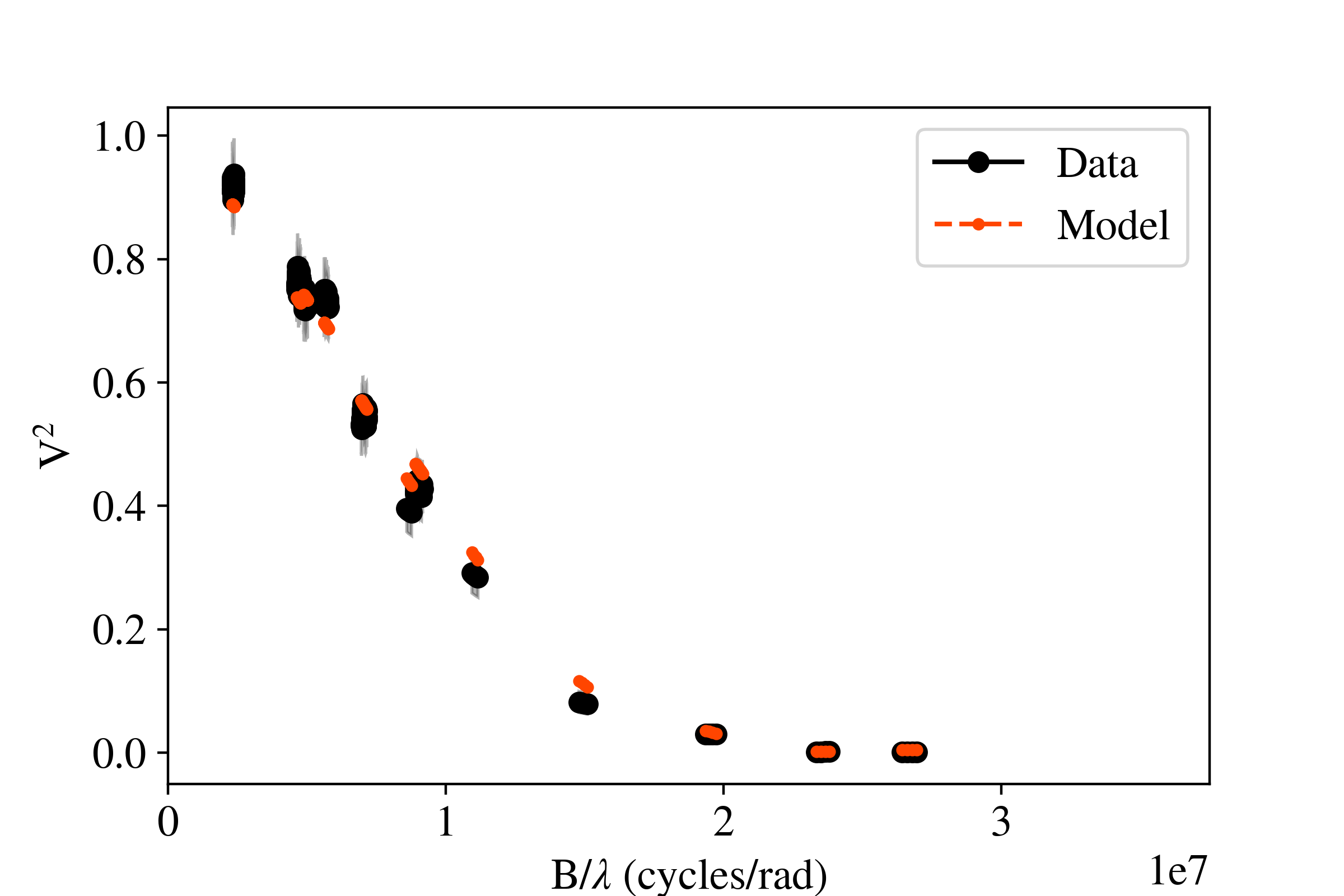}
    \includegraphics[height=0.24\textwidth]{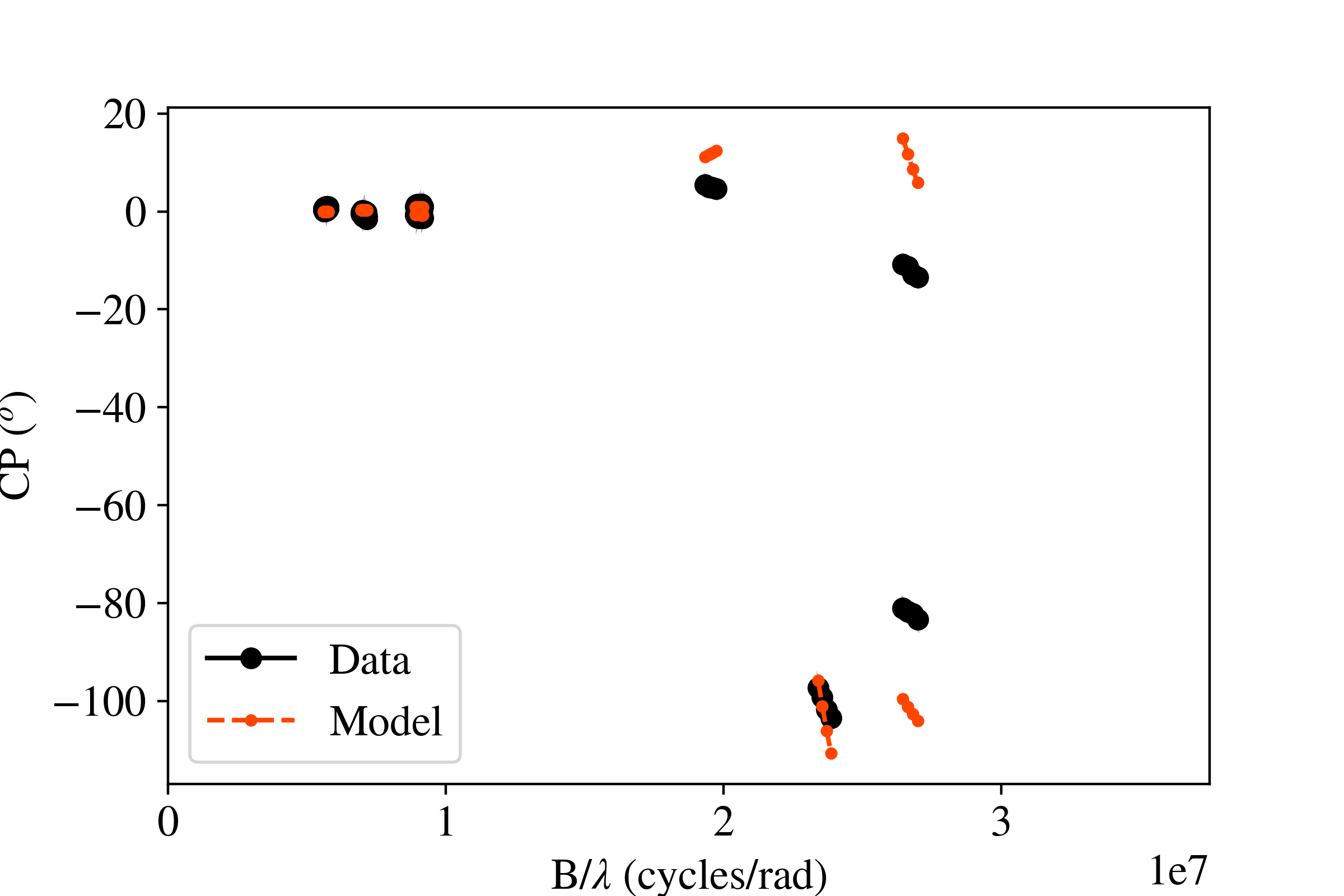}
    \includegraphics[width=0.24\textwidth]{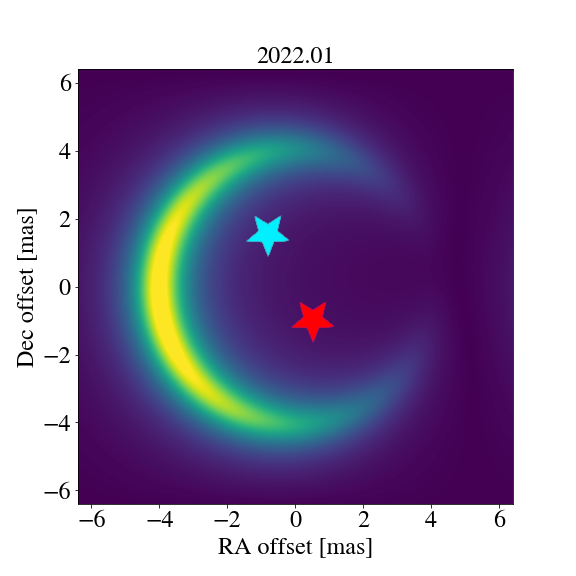}
    \includegraphics[height=0.24\textwidth]{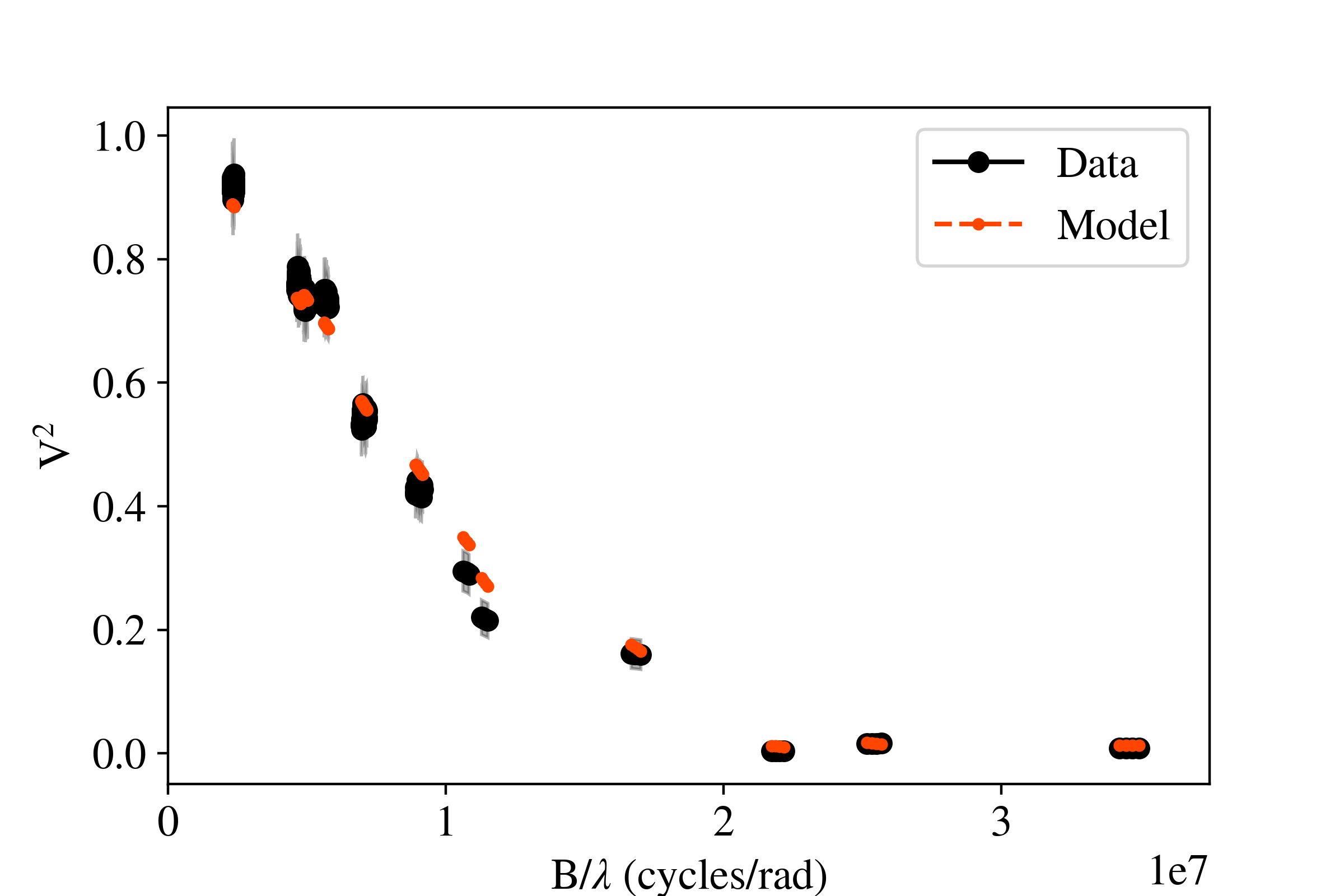}
    \includegraphics[height=0.24\textwidth]{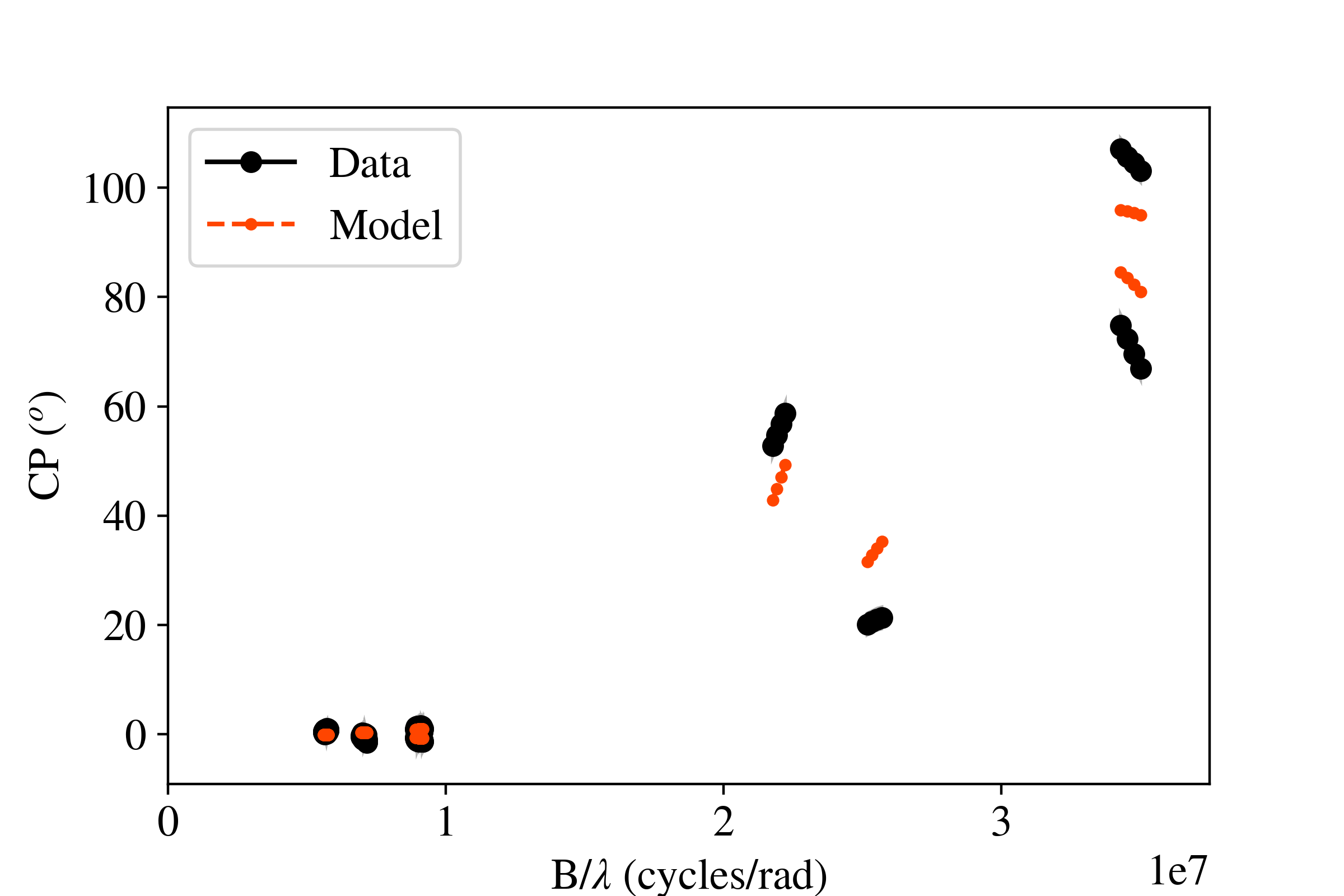}
    \caption{Results of the model fitting at the 2020, 2021, and 2022 epochs (we also overplotted the 2023 data in all cases). First column: Model images. The primary is marked by a blue star symbol, the secondary by a red star symbol. Second column: Fitting of the visibilities. The data marked by black symbols, the model marked by red symbols. Third column: Fitting of the closure phases. The marking is similar to the previous column.}
    \label{fig:images}
\end{figure*}

The relative flux contributions of the components are important parameters in our disk model. To estimate the flux ratios of the stars at the wavelengths of our MATISSE data, we performed an SED analysis. 
The temperatures of the stars are taken from \cite{2013MNRAS.431.3485C}, with $T_\mathrm{eff}=8250$ K for the primary, and $T_\mathrm{eff}=4800$ K for the secondary. To reproduce the stellar emission, we used PHOENIX synthetic stellar model atmosphere spectra by \cite{2013A&A...553A...6H}. For the primary, we took a model spectrum with $T_\mathrm{eff}=8200$ K, log $g=+3.5$, and [Fe/H]$=0.0$. For the secondary, we used a spectrum with $T_\mathrm{eff}=4800$ K, log $g=+4.0$, and [Fe/H]$=0.0$. The flux ratio of the secondary vs. primary at $557.6$~nm, $1/9.9$, is taken from \cite{2013MNRAS.431.3485C}, and the model spectra were scaled accordingly\footnote{For the scaling, we used the average of the PHOENIX spectra in a $100$ nm wide window centered on $557.6$~nm. 
The properly scaled model spectra were then added together, and fitted to photometric data using the \texttt{SED Fitter} tool \citep{2007ApJS..169..328R}. For the photometry, we used Gaia $G$, $G_\mathrm{BP}$, $G_\mathrm{RP}$ \citep{2018A&A...616A...1G}, Hipparcos $BT$, $VT$ \citep{hog2000}, and 2MASS $J$ \citep{cutri2003} magnitudes. The interstellar extinction was taken into account in the fitting, with $A_V=0.31$ \citep{vizier2018}.} 
The results are shown in Figure \ref{fig:sed}.

To determine the $L$-band flux contributions of each component, we used the total (single-dish) spectra of the four MATISSE measurements, representative of the emission of the entire system, and the properly scaled stellar model atmosphere spectra. For each MATISSE measurement, we calculated the average of the fluxes between $3.45$ and $3.55$ $\upmu$m. For both stellar model spectra, we also averaged the fluxes in the same wavelength range. The resulting flux ratios for each epoch are shown in Table \ref{tab:flux}.

\begin{table}
\caption{Flux contributions of the DX Cha system elements. The percent values are rounded to the nearest hundreds.}
\begin{center}
    \label{tab:flux}
    \small
    \begin{tabular}{c c c c}
        \hline
        \hline
         & \multicolumn{3}{c}{Flux contribution [\%]}\\
        \hline
        Epoch & Disk & Primary & Secondary\\
         &  & star & star\\
        \hline
        2020 & 85.45 & 8.89 & 5.65\\
        2021 & 86.68 & 8.14 & 5.18\\
        2022 & 84.57 & 9.42 & 6.00\\
        2023 & 89.74 & 6.27 & 3.99\\
        \hline
         
    \end{tabular}
\end{center}
\end{table}

To determine the structure and properties of the circumbinary disk of DX Cha, we built a customized geometric model using the \texttt{oimodeler}\footnote{\url{https://github.com/oimodeler/oimodeler}, version 0.85-beta} modeling software package \citep{Meilland2024_oimodeler}.

We used an azimuthally modulated ring to reproduce the disk, based on the model of \cite{Lazareff2017}. The ring has a diameter, convolved with a Gaussian kernel. The full width at half maximum (FWHM) of the kernel determines the width of the ring. The diameter ($d$) and the FWHM are fitted parameters, but they have the same values in all epochs. The azimuthal modulation is of first order, which means that there is one brightness maximum and, in the opposite direction, one brightness minimum, over the whole azimuthal range, and the azimuthal brightness intensity variation follows a cosine shape. Since we aim to study the time variability of asymmetry, we fit the amplitude of the azimuthal modulation separately for the three epochs with long baseline measurements ($A_{mod1}$, $A_{mod2}$, $A_{mod3}$). Similarly, we fit the position angle of the azimuthal modulation separately for the three epochs ($\Phi_{mod1}$, $\Phi_{mod2}$, $\Phi_{mod3}$).
All earlier observations showed that the inclination of the DX Cha circumbinary disk is close to face-on (e.g. $i=18^{\circ}$, \citealp{2004ApJ...608..809G}), thus we used a face-on disk in our model to avoid degeneracies arising from the coupling of $i$, $\Omega$, and $\Phi$.

To reproduce the binary, we placed two point sources with different masses and luminosities inside the asymmetric ring. In our model, the binary components follow Keplerian (2-body) motion, and its orbital element $t_0$ (time of periastron passage) is a fitted parameter. Although \cite{2004A&A...427..907B} already fitted the time of the periastron passage, its error is unknown. 
Even a small inaccuracy in $t_0$ can lead to a large error in the orbital phase by the time of our MATISSE measurements. Thus, we do not use the value reported by \cite{2004A&A...427..907B}, instead, we fit $t_0$ as a free parameter.

The other orbital parameters are fixed to the values in Table \ref{tab:fix}. All stellar parameters we used based on the literature are listed in Table \ref{tab:fix}. We used the python package \texttt{twobody}\footnote{\url{https://github.com/adrn/TwoBody}, version 0.8.3.} to calculate the positions of the stars. Although the disk was modeled face-on, we use an inclined binary orbital plane in our model, in accordance with the literature (Table \ref{tab:fix}). The center of the ring is fixed at the barycenter. We fit the flux contributions of the primary and the secondary star ($f_p$, $f_s$) in narrow ranges based on our SED analysis. The fit ranges are set by the minimum and maximum values of the per-epoch flux ratios reported in Table~\ref{tab:flux} (separately for the primary and secondary).

We fit the MATISSE squared visibility and closure phase data in the $3.45$--$3.55\ \upmu$m wavelength range, simultaneously for all three epochs. Such a small wavelength range helps to avoid chromatic effects (e.g., the wavelength-dependence of the size of the disk emitting region). As the error bars on the MATISSE data products might be underestimated \citep[e.g.,][]{varga2021}. Based on the error analysis of \cite{varga2021} we imposed a minimum error of $0.03$ on the squared visibility, and $2.0^\circ$ on the closure phase. For the parameter search, we used \texttt{emcee} \citep{2013PASP..125..306F}, a Markov Chain Monte Carlo (MCMC) ensemble sampler with 50,000 steps and 26 walkers, with a burn-in of 25,000 iterations. Our best-fit values (Table \ref{tab:bestfit}) are the ones with the lowest $\chi^{2}_{red}$, which is calculated in the following way:

\begin{equation}
\chi^{2}_{red}=\frac{1}{\nu} \Sigma_{i} \frac{\left(O_{i}-C_{i}\right)^2}{\sigma_{i}^2}
\end{equation}

where $O_{i}$ are the data points, $C_{i}$ are the model values, and $\sigma_{i}$ are the uncertainties of the data.
The degree of freedom, $\nu=n-m $, equals the number of data points $n$ minus the number of fitted parameters $m$. The errors on the parameters are calculated as the $16$--$84$ percentile ranges of the resulting posterior distributions (Appendix \ref{sec:app_modelplots}). The squared visibility and closure phase data was equally weighted during the calculations.

\section{Results}
\label{sec:results}

\begin{table*}
\caption{The best-fit parameters of the modeling, with a $\chi^{2}_{red}=1.13$. Position angles of the modulations are measured east of north.}
\begin{center}
    \label{tab:bestfit}
    \small
    \begin{tabular}{c c c c c c c c c c c}
        \hline
        \hline
         $d$ & $FWHM$ & $A_{mod1}$ & $A_{mod2}$ & $A_{mod3}$ & $\Phi_{mod1}$ & $\Phi_{mod2}$ & $\Phi_{mod3}$ & $t_0$ & $f_p$ & $f_s$ \\
         
         [mas] & [mas] & [$^{\circ}$] & [$^{\circ}$] & [$^{\circ}$] & [$^{\circ}$] & [$^{\circ}$] & [$^{\circ}$] & [MJD] & \% & \%\\
        \hline
$8.10^{+0.49}_{-0.42}$ & $1.23.^{+0.24}_{-0.46}$ & $0.77^{+0.39}_{-0.25}$ & $0.47^{+0.14}_{-0.23}$ & $0.98^{+0.40}_{-0.22}$ & $109.01^{+42.65}_{-20.29}$ & $253.35^{+31.61}_{-51.47}$ & $262.45^{+25.06}_{-47.21}$ & $58922.01^{+2.88}_{-2.65}$ & $9.30^{+2.70}_{-1.60}$ & $6.00^{+1.50}_{-0.80}$ \\
        \hline
    \end{tabular}
\end{center}
\end{table*}

The best-fit model images for each epoch, along with the fits to the data are shown in Figure \ref{fig:images}. 
The best-fit model parameters are listed in Table \ref{tab:bestfit}. The posterior distributions of the MCMC samples are shown in the Appendix \ref{sec:app_modelplots}.

Our model can fit the data well in all three epochs. For all epochs, taking into account squared visibilities and closure phases $\chi^{2}_{red}=1.13$. The $\chi^{2}_{red}$ values are shown separately for the squared visibilities and closure phases, as well as for each epoch in Table \ref{tab:chired}. Apart from the closure phases in epoch 2 and epoch 3, the $\chi^{2}_{red}$ values are below one, indicating excellent correspondence between data and model.

\begin{table}
\caption{The $\chi^{2}_{red}$ values are for the squared visibilities and closure phases at the each epochs. $\star$In all calculations, we also took into account the 2023, small baseline measurements.}
\begin{center}
    \label{tab:chired}
    \small
    \begin{tabular}{c c c c}
        \hline
        \hline
         & \multicolumn{3}{c}{$\chi^{2}_{red}$} \\
        \hline
        Epoch$\star$ & Squared visibility & Closure phase & Squared visibility \\
         &  &  & and \\
         &  &  & Closure phase \\
        \hline
        2020 & 0.54 & 0.51 & 0.52\\
        2021 & 0.46 & 3.80 & 1.76\\
        2022 & 0.48 & 2.14 & 1.13\\
        \hline
    \end{tabular}
\end{center}
\end{table}

The values of the size parameters $d$ and $FWHM$ show that a  narrow ring with a diameter of $\approx$$0.86$~au is detectable in the $L$-band. 
The fitted values of $A_{mod1}$, $A_{mod2}$, $A_{mod3}$, representing the amplitude of the asymmetry, indicate a significant asymmetry in the brightness distribution of the ring. The posteriors show (Appendix \ref{sec:app_modelplots}) that $A_{mod1}$ and $A_{mod3}$ strongly favor a value of 1.0 which is the largest permitted value. Modulation amplitudes larger than that are unphysical because they would imply negative surface brightness values. The preference for a large asymmetry amplitude may indicate that the azimuthal brightness profile of the disk asymmetry is not fully reproduced by our model. Nevertheless, the fits to the data are satisfactory, therefore we refrain from adding a more complex geometry for the azimuthal asymmetry. We find that the position angle of the asymmetry is time-variable, as evidenced from the significantly different values of $\Phi_{mod1}$, $\Phi_{mod2}$, and $\Phi_{mod3}$.
The best-fit model images show that in the three MATISSE epochs the stars are aligned approximately in the north-south direction, with a relatively large separation ($\sim$$0.3$~au).

\section{Discussion}
\label{sec:disc}

The best-fit amplitude of the modulation parameters of the ring are markedly different from zero, which also confirms the asymmetry of the innermost disk region. The direction of this asymmetry also changes over time. However, from the current data alone, we cannot make a statement regarding the periodicity of the changes.

\begin{figure}[!ht]
\centering
    \includegraphics[width=0.95\columnwidth]{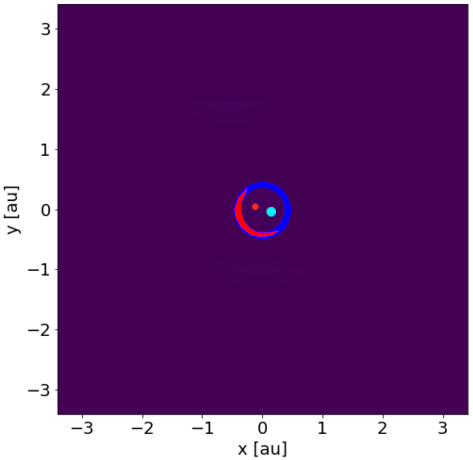}
    \includegraphics[width=0.95\columnwidth]{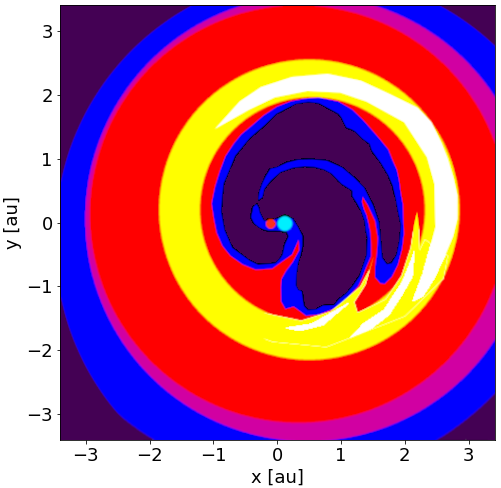}
    \caption{Schematic representations of models of DX Cha, for the comparison of the relevant disk structures, without exact color scales. Top: our model, based on the $L$-band brightness distribution. Bottom: sketch of the surface density map from the SPH simulation by \cite{2015MNRAS.448.3545D}. The epoch of the MATISSE $L$-band model was chosen to show the stars in an alignment similar to that in the SPH simulation snapshot. The blue dot indicates the primary, and the red dot the secondary.}
    \label{fig:dunhill_and_model}
\end{figure}

The SPH simulation of \cite{2015MNRAS.448.3545D} was based on the stellar parameters of DX Cha, and they noticed that the eccentric binary carves a large inner cavity with a diameter of $\sim$$4$~au. In contrast, the circumbinary ring in our model is much smaller, only $0.86$ au in diameter. Thus, we argue that the $L$-band emitting structure seen in our data is not the cavity inner wall. To illustrate the case, in Figure \ref{fig:dunhill_and_model} we compare a snapshot of the SPH simulation \citep{2015MNRAS.448.3545D} with our model on the same spatial scale. The model image shows a situation where the stars are located roughly along the east-west axis, while the ring geometry corresponds to the 2021 epoch.

\cite{2015MNRAS.448.3545D} predicted that a circumbinary disk should precess around the stars with a period of approximately 40 years, leading to noticeable changes within just a few years. However, we do not observe this in our data, likely because we do not detect the inner wall of the disk, where variations may occur on shorter timescales corresponding to the binary period. Therefore, advances in high-cadence (weekly to monthly) and long-term (spanning multiple years) monitoring observations with VLTI, preferably at longer wavelengths, would be needed to detect the disk wall.

The surface density maps of the SPH simulation show spiral-like accretion bridges within the cavity. We propose that the $L$-band dust emission, represented as a ring in our model, may correspond to the inner part of those accretion bridges, close to the dust sublimation radius, that is warm and dense enough to emit significant mid-IR emission. The expected dust sublimation radius for the primary is $\sim$$0.4$~au, which is similar to the ring radius in our model. We note that \cite{2015MNRAS.448.3545D} plotted surface density values, while we model the brightness distribution. Thus, a direct comparison between the two models is not possible without calculating the radiative transfer on the SPH density map.
  
Based on our modeling, it can be established that the Keplerian orbit assumed for the binary is consistent with the observational data. 
The results indicate that the binary star is at a large separation and at a similar orientation in all three epochs. If we consider the orbital period of $P=19.859 d$ by \citep{2004A&A...427..907B}, and we denote the Julian dates of the 2020, 2021, and 2022 observations as $JD1$, $JD2$, and $JD3$, respectively, we get
\begin{equation}
(JD2-JD1)/P = 17.82
\end{equation}
\begin{equation}
(JD3-JD2)/P = 16.02
\end{equation}
This means that epoch 2 and epoch 3 has almost the same orbital phase because a close to integer number of full orbits elapsed between them. The situation is similar for epoch 1 and epoch 2.
The same comparison can be done between our epochs and the epoch of periastron from \citet{2004A&A...427..907B} ($JD0 = 2451647.505$, $MJD0 = 51647.005$). Then we get the following:
\begin{equation}
(JD1-JD0)/P = 366.75
\end{equation}
\begin{equation}
(JD2-JD0)/P = 384.57
\end{equation}
\begin{equation}
(JD3-JD0)/P = 400.59
\end{equation}
This shows that at least JD2 and JD3 are close to apastron (which would be phase 0.5). All of these suggest a good agreement between the original spectroscopic monitoring results of \citet{2004A&A...427..907B} and our new modeling results from MATISSE data.

Previous $K$-band studies \citep{2007A&A...464...55T, Kraus2008, 2013MNRAS.430.1839G} also found material in the $0.2$--$0.5$ au radial region. It can be assumed that they detected the same structures in the $K$-band AMBER data as we did in our $L$-band analysis.

\section{Summary}
\label{sec:sum}

In this work, we studied the structure of the disk around the spectroscopic binary DX Cha, based on VLTI/MATISSE $L$-band interferometric observations taken at four epochs between 2020 and 2023. Using the \texttt{oimodeler} software, we constructed an interferometric model of the source, consisting of an asymmetric skewed ring, and two point sources on a Keplerian orbit. Our main results are summarized in the following.
\begin{itemize}
    \item Our model indicates that the $L$-band circumbinary emission is concentrated in a narrow ring with a diameter of $8.1$ mas ($0.86$~au at 106.5\,pc).
    \item The best-fit values of $A_{mod1}$, $A_{mod2}$, $A_{mod3}$ ($0.77^{+0.39}_{-0.25}$, $0.47^{+0.14}_{-0.23}$, $0.98^{+0.40}_{-0.22}$) show a significant azimuthal asymmetry of the ring.
    Moreover, we found that the position angle of the asymmetry is significantly different in the three epochs, indicating that the position of the brightness peak changes over time. The periodicity of the asymmetry cannot be constrained from the current data set. 
    \item We demonstrated that both the binary star and the asymmetrically distributed circumstellar material are significant contributors to the observed closure phase signal.
    \item Earlier hydrodynamical simulations predicted that the DX Cha binary would carve a large cavity in the inner disk with a diameter of $\sim$$4$~au. However, we found that the entire $L$-band emitting structure is well inside the assumed cavity. Thus, we conclude that we do not detect the cavity inner wall in our data. The $L$-band emitting structure may be the inner warm part of the accretion bridges channeling material from the disk toward the stars.
    
\end{itemize}

\newpage
\bibliography{bibliog}{}
\bibliographystyle{aasjournal}

\begin{acknowledgements}
MATISSE was designed, funded and built in close collaboration with ESO, by a consortium composed of institutes in France (J.-L. Lagrange Laboratory -- INSU-CNRS -- C\^ote d’Azur Observatory -- University of C\^ote d'Azur), Germany (MPIA, MPIfR and University of Kiel), the Netherlands (NOVA and University of Leiden), and Austria (University of Vienna). The Konkoly Observatory and Cologne University have also provided some support in the manufacture of the instrument.

Based on observations collected at the European Southern Observatory under ESO programmes 190.C-0963(D), 190.C-0963(E), 190.C-0963(F), 0100.C-0278(E), 0103.D-0153(C), 0103.C-0347(C), 0103.D-0153(G), 108.225V.003, 108.225V.011, and 108.225V.006.

The project was supported by the Hungarian OTKA grants K132406 and K147380.

F. Cruz-Sáenz de Miera received financial support from the European
Research Council (ERC) under the European Union’s Horizon 2020 research and innovation programme (ERC Starting Grant ``Chemtrip", grant agreement
No 949278).

J. Varga acknowledges support from the Fizeau exchange visitors program. The research leading to these results has received funding from the European Union’s Horizon 2020 research and innovation programme under Grant Agreement 101004719 (ORP).

The open-source \texttt{Oimodeler} package used in this research is developed with support from the VLTI/MATISSE consortium and the ANR project MASSIF, and we would like to thank the whole development team with an explicit mention to Anthony Meilland, Marten Scheuck and Alexis Matter for their work.

This research has made use of the services of the ESO Science Archive Facility.

This research has made use of the VizieR catalog access tool, CDS, Strasbourg, France (DOI : 10.26093/cds/vizier). The original description of the VizieR service was published in 2000, A\&AS 143, 23.
 
This work has made use of data from the European Space Agency (ESA) mission
{\it Gaia} (\url{https://www.cosmos.esa.int/gaia}), processed by the {\it Gaia}
Data Processing and Analysis Consortium (DPAC,
\url{https://www.cosmos.esa.int/web/gaia/dpac/consortium}). Funding for the DPAC
has been provided by national institutions, in particular the institutions participating in the {\it Gaia} Multilateral Agreement.
This study has made use of the \textsc{GGchem} code, available at \url{https://github.com/pw31/GGchem}.
\end{acknowledgements}

\newpage
\begin{appendix}
    
\section{Supplementary model plot}
\label{sec:app_modelplots}
\begin{figure*}[!ht]
\centering
    \includegraphics[width=\textwidth]{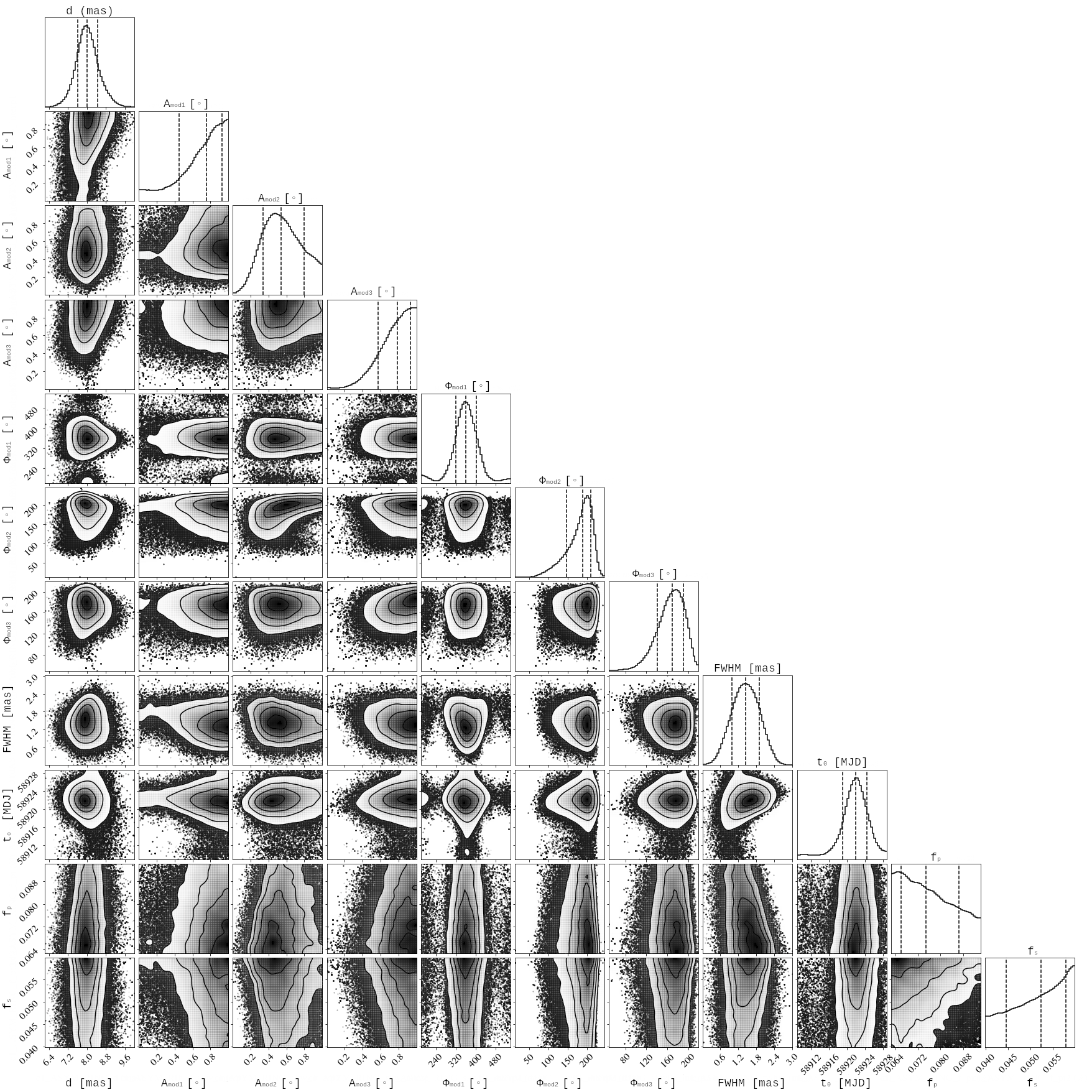}
    \caption{Corner plot showing the posterior distributions of the MCMC samples.}
    \label{fig:corner}
\end{figure*}

\end{appendix}

\end{document}